\documentclass[twocolumn]{aastex631}
\usepackage{CJKutf8} 
\usepackage{amsmath}
\usepackage{comment}
\usepackage{physics}

\shortauthors{Goto et al.}
\shorttitle{Ultra-diffuse Satellites}

\begin{document}

\title{Systematically Measuring Ultra-Diffuse Galaxies (SMUDGes). IV.  Ultra-Diffuse Satellites of Milky Way Analogs}
  

\author[0000-0002-0268-5707]{Hina Goto}
\affiliation{Steward Observatory and Department of Astronomy, University of Arizona, 933 N. Cherry Ave., Tucson, AZ 85721, USA}

\author[0000-0002-5177-727X]{Dennis Zaritsky}
\affiliation{Steward Observatory and Department of Astronomy, University of Arizona, 933 N. Cherry Ave., Tucson, AZ 85721, USA}

\author[0000-0001-8855-3635]{Ananthan Karunakaran}
\affiliation{Instituto de Astrof\'{i}sica de Andaluc\'{i}a (CSIC), Glorieta de la Astronom\'{i}a, 18008 Granada, Spain}

\author[0000-0001-7618-8212]{Richard Donnerstein }
\affiliation{Steward Observatory and Department of Astronomy, University of Arizona, 933 N. Cherry Ave., Tucson, AZ 85721, USA}

\author[0000-0003-4102-380X]{David J. Sand }
\affiliation{Steward Observatory and Department of Astronomy, University of Arizona, 933 N. Cherry Ave., Tucson, AZ 85721, USA}

\begin{abstract}
To better understand the formation of large, low surface brightness galaxies, we measure the correlation function between ultra-diffuse galaxy (UDG) candidates and Milky Way analogs (MWAs). We find that (1) the projected radial distribution of UDG satellites (projected surface density $\propto r^{-0.84\pm0.06}$) is consistent with that of normal satellite galaxies, (2) the number of UDG satellites per MWA ($S_{\rm UDG}$) is $\sim 0.5\pm0.1$ over projected radii from 20 to 250 kpc and $-17< M_r < -13.5$, (3) $S_{\rm UDG}$ is consistent with a linear extrapolation of the relationship between the number of UDGs per halo vs. halo mass obtained over galaxy group and cluster scales, (4) red UDG satellites dominate the population of UDG satellites ($\sim80$\%), 
(5) over the range of satellite magnitudes studied, UDG satellites comprise $\sim$ 10\% of the satellite galaxy population of MWAs, (6) a significant fraction of these ($\sim$13\%) have estimated total masses $>$ 10$^{10.9}$ M$_\odot$ or, equivalently, at least  half the halo mass of the LMC, and populate a large fraction ($\sim$ 18\%) of the expected subhalos down to these masses. All of these results suggest a close association between the overall low mass galaxy population and UDGs, which we interpret as favoring models where UDG formation principally occurs within the general context of low mass galaxy formation over models invoking more exotic physical processes specifically invoked to form UDGs. 

\end{abstract}


\section{Introduction}
\label{sec:intro}

The success of the $\Lambda$CDM paradigm as a predictive framework for structure formation is nearly complete, with the only unresolved issues remaining at small galaxy ($\ll$ L$^*$) scales \citep[for an overview see][]{weinberg}. The study of low mass galaxies is then expected to highlight important baryonic physical evolutionary processes that may be missing in the simulations and, perhaps even more excitingly, potential departures from the canonical CDM phenomenology. The desire for progress on either of these two fronts has motivated significant efforts to improve the empirical study of low mass galaxy populations \citep[such as the deep galaxy searches undertaken within a variety of nearby environments;][]{sand,park1,venhola17,virgo,lamarca}. 

A particular class of low mass galaxy is that of the satellite galaxy. Satellites lie within an even larger halo of a parent or host system. We focus here on satellite galaxies lying within the halos of $\sim$ L$^*$ galaxies that we refer to as Milky Way analogs (MWAs). Due to gravitational and  hydrodynamical interactions with these parent galaxies, simulations suggest that the numbers, internal structure, and star formation histories of such satellites may have been altered relative to what they would have been for the same galaxies in isolation  \citep[for recent examples of such work see;][]{martin,samuel}. As such, there are a variety of reasons to compare samples of satellite galaxies to samples of similar mass galaxies that do not consist exclusively of satellites \citep[e.g., that of ][]{blanton}. 

Broadly, there are two approaches used to search for satellite galaxies. In the first, which we refer to as ``photometric", imaging is used to identify potential satellites around a set of selected parent galaxies. Typically, the redshifts are known for the parent galaxies but not for the candidate satellites. One measures the bulk satellite properties by evaluating the excess population of candidates projected in the vicinity of the parent galaxies \citep{holmberg,lorrimer,sales05,guo,wang12,sales}.
Because foreground and background objects generally greatly outnumber satellites, large samples are needed to tease out results. In this approach one is able, for example, to reach conclusions regarding the radial density profile of satellites around parents, but not to determine which of the many candidates are true satellites. With the advent of large area photometric surveys, this approach now brings statistical power to the questions at hand and subsamples can be defined to explore properties of the satellite galaxy population. In the second approach, which we refer to as ``spectroscopic", one measures redshifts of the satellite candidates and identifies those sharing the parent's redshift to within allowances for different peculiar velocities \citep{z93,prada,geha-saga}. This approach does then allow for follow-up of the true satellites and for an examination of the satellite kinematics but comes at great observational expense because it requires spectroscopy of many faint targets, most of which are contaminants. As such, it currently provides lower precision on the determination of certain bulk properties of satellites, such as the radial density profile, and is, of course, limited to satellites that are within spectroscopic reach.

Both of these approaches are limited by the initial selection of the candidate satellites, which is based on imaging and will always suffer from a surface brightness selection effect (spectroscopic surveys suffer an additional surface brightness bias because of the further difficulty in obtaining the spectra). The recent appreciation that there are many fairly large galaxies, large both in physical size (effective radius, $r_e$, $> 1.5$ kpc) and total luminosity (some as bright as $M_g \sim -17$), that have evaded detection due to their exceedingly low central surface brightness, and that such galaxies survive in dense environments \citep{vdk15a}, leads to a suspicion that previous satellite samples may be missing satellites that are as massive as the Large Magellanic Cloud \citep[e.g., DF 44 has M$_{200} = 10^{11.2\pm0.6}$ M$_\odot$;][]{vdk19df44}. 

This suspicion prompted the recent examination of the deepest available samples of satellite galaxies outside the Local Group \citep{carlsten,2021Mao,2022Nashimoto} by \cite{karunakaran22} for what those samples have to say regarding the existence of low surface brightness, physically-large galaxies (commonly referred to as ultra-diffuse galaxies or UDGs) as satellites of L$^*$, or MWA, galaxies. Their conclusion was that MWAs host proportionally, by total mass, nearly the same number of UDGs as do more massive halos. 

On its surface, this result suggests that UDG formation is neither enhanced nor inhibited in the galactic environment.   Nevertheless, the sample of UDG satellites in that study consisted of only 41 confirmed\footnote{Seventeen confirmed  with spectroscopic redshifts and 24 with distances measured using surface brightness fluctuations.} satellites split among 75 parent galaxies, making it difficult to divide the sample into categories and address additional questions. Encouragingly, consistent conclusions regarding the mean number of such  satellites for MWAs were presented by \cite{li},
who use an enhanced photometric approach that incorporates size and color to help remove contamination from a sample without spectroscopic follow-up. It is important to note that for most UDGs a spectroscopic approach is not feasible given that exposure times of $\sim$ 1 hour are needed on large telescopes to obtain a redshift \citep{vdk15b,chilingarian,kadowaki21}.

We return to the photometric approach with a  
large sample of UDG candidates and focus on bulk UDG satellite properties. We undertake this study because there now exists a catalog of UDG candidates that spans nearly 20,000 sq. degree of sky and contains nearly 7,000 candidates \citep[the SMUDGes catalog;][]{smudges,smudges3,smudges4}. We aim to establish the radial number density profile, the luminosity function, and the color distribution of UDG satellites of MWAs and compare those to the corresponding measurements of the more classical satellite population. By doing so, we will present conclusions regarding plausible UDG formation and evolution models. We present the technical aspects of the approach in \S\ref{sec:method} and our results and interpretation in  \S\ref{sec:results}. 
We use a standard WMAP9 cosmology \citep{wmap9}, although the results are insensitive to different choices of cosmological parameters at the level of current uncertainties, and magnitudes are from SDSS/DESI and are thus on the AB system \citep{oke1,oke2}.

\section{Methodology}
\label{sec:method}

We use the SIMBAD database \citep{simbad} to identify Milky Way analogs (MWAs) projected in proximity to each UDG candidate in our catalog that meets a minimum 20\% estimated completeness  \cite[for details of the completeness calculation see][but this criterion corresponds to an estimate that we have found $>$ 20\% of the UDG candidates with similar photometric properties across the full survey footprint]{smudges3}. We impose this cut to avoid having to make large, highly uncertain completeness corrections. Overall the completeness is roughly 50\%, due mainly to aggressive masking of the survey area around bright foreground objects and regions of Galactic Cirrus. An important related concern is that the completeness is expected to fall dramatically near each MWA because those regions were masked in the original UDG search \citep{smudges3}. That incompleteness factor is included in an average sense in the catalog because we account for the masked area but is not mapped specifically around each MWA. As such, the distribution of pair separations will be increasingly incorrect at ever smaller separations, but, for the most part,  we work at projected radii where we do not expect this to have an impact. Nevertheless, we search for signs of this effect in the results discussed below.

Using the absolute magnitude of the Milky Way \citep[M$_g = -21.0$;][]{bland-hawthorn}, we define MWAs to have $-22 < {\rm M}_g < -20$ in three different recessional velocity slices ($4500<cz/({\rm km\ s}^{-1})<5500$, $5500<cz/({\rm km\ s}^{-1})<6500$, and $6500<cz/({\rm km\ s}^{-1})<7500$). The three slices provide us with independent checks of the results. Caution is warranted when comparing among results from different studies, as the definition of MWA varies among studies focusing on such objects \citep[e.g.,][]{2021Mao,carlsten}, which often have additional environmental conditions or, perhaps less critically, slightly different magnitude criteria. 
We impose the lower $cz$ limits on the parents to ensure that candidate UDGs, which are selected in SMUDGes to have $r_e > 5.3$ arcsec, match the physical size criterion of UDGs \citep[$r_e \ge $ 1.5 kpc;][]{vdk15a}. We limit the redshift range of each slice to minimize possible variations in satellite magnitude and size within each sample. We adopt an upper size limit ($r_e < 6$ kpc), which is not a standard UDG criterion, because UDG candidates with inferred sizes larger than this limit are likely contaminants \citep{kadowaki21,smudges3,karunakaran22}.
To include potential MWAs without a cataloged value of m$_g$ that do have a cataloged value of m$_B$, we calculate the average m$_g - $m$_B$ color for those with photometry in both bands and apply that value as a correction to m$_B$ for those missing m$_g$. This is a crude correction but we only use the parent magnitude to place it relative to the two magnitude wide bin defining MWAs. 

\begin{figure}[ht]
    \centering
    \includegraphics[scale=0.5]{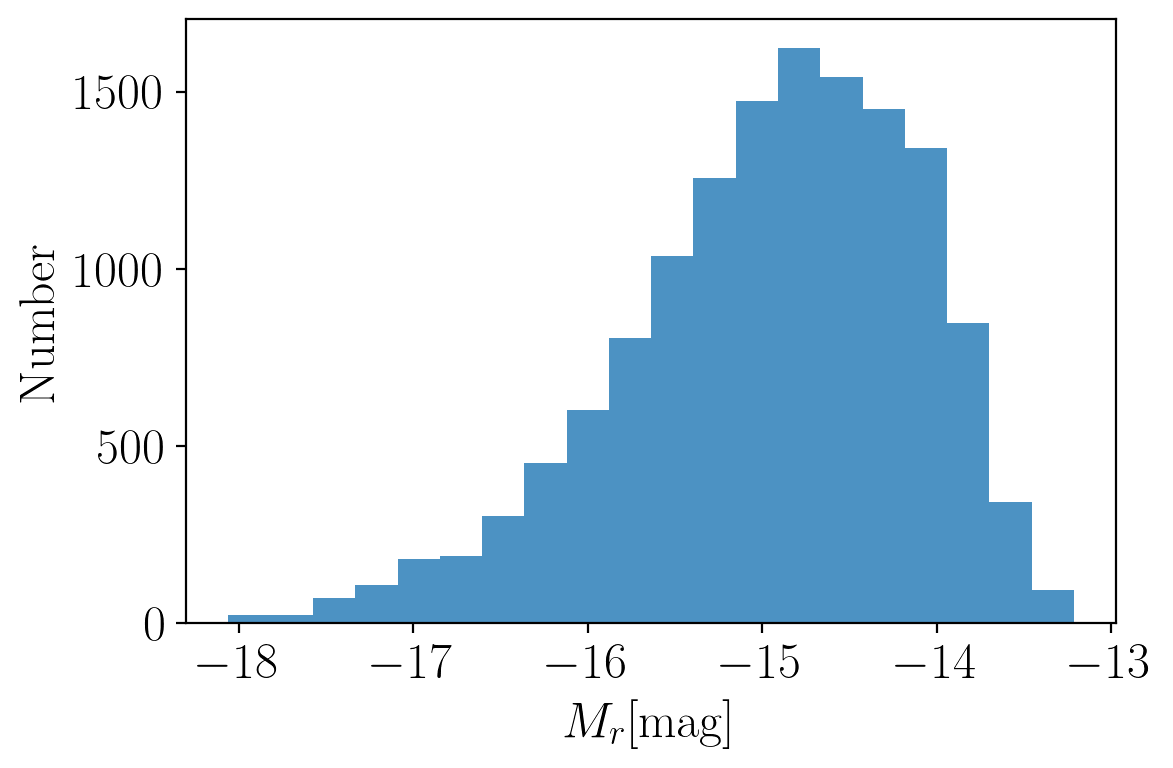}
    \caption{{Distribution of inferred absolute magnitudes of UDG satellites when working within the $4500 < cz$/(km sec$^{-1}) < 5500$ slice. Although many of the galaxies are likely to be false pairs, this plot shows the type of UDG satellites to which we are sensitive.}}
\label{fig:mags}
\end{figure}

We set the search radius for MWAs around each UDG candidate in
the completed SMUDGes catalog \citep{smudges4} to correspond to 10 Mpc at the near edge of each recessional velocity slice because that separation corresponds roughly to where the galaxy-galaxy correlation function drops below a value of 1 \citep{tucker,zehavi}.
From each SIMBAD query, we retain the right ascension ($\alpha$), declination ($\delta$), m$_g$, m$_B$, and redshift of each MWA candidate.
We assign the UDG candidate the redshift of the associated MWA, evaluate its absolute magnitude and size, reject candidates that do not match our physical size criteria for UDGs, and calculate the projected physical separation between the pair. In Figure \ref{fig:mags} we present the inferred absolute magnitude distribution of the candidate UDGs. Although the majority of these values are drawn from unphysical pairs, and are therefore incorrect, the distribution does provide some intuition regarding the types of satellites to which this analysis is sensitive.

Our search produces a list of 626,560, 411,507, and 538,726 accepted UDG-MWA pairs for the three different redshift slices from the catalog of 6332 UDG candidates and the recovered 2905 MWAs across the three redshift slices. We track the distribution of pair separations, inversely weighted by the completeness fraction of each UDG separately for (1) all pairs, (2) those containing a blue UDG, and (3) those containing a red UDG. The color dividing line is defined to be 0.1 mag bluer than the empirically-determined red sequence for UDGs \cite[$g-r = 0.167 -0.031 M_r$;][]
{smudges3}. We also set lower ($g-r > 0.2 $) and upper ($g-r < 0.75 $) color cutoffs to remove likely contaminants in the catalog \citep{smudges3}. 
The blue limiting criterion also matches the color of the bluest field UDGs examined in detail \citep{jones23}.

\begin{figure}[ht]
    \centering
    \includegraphics[scale=0.5]{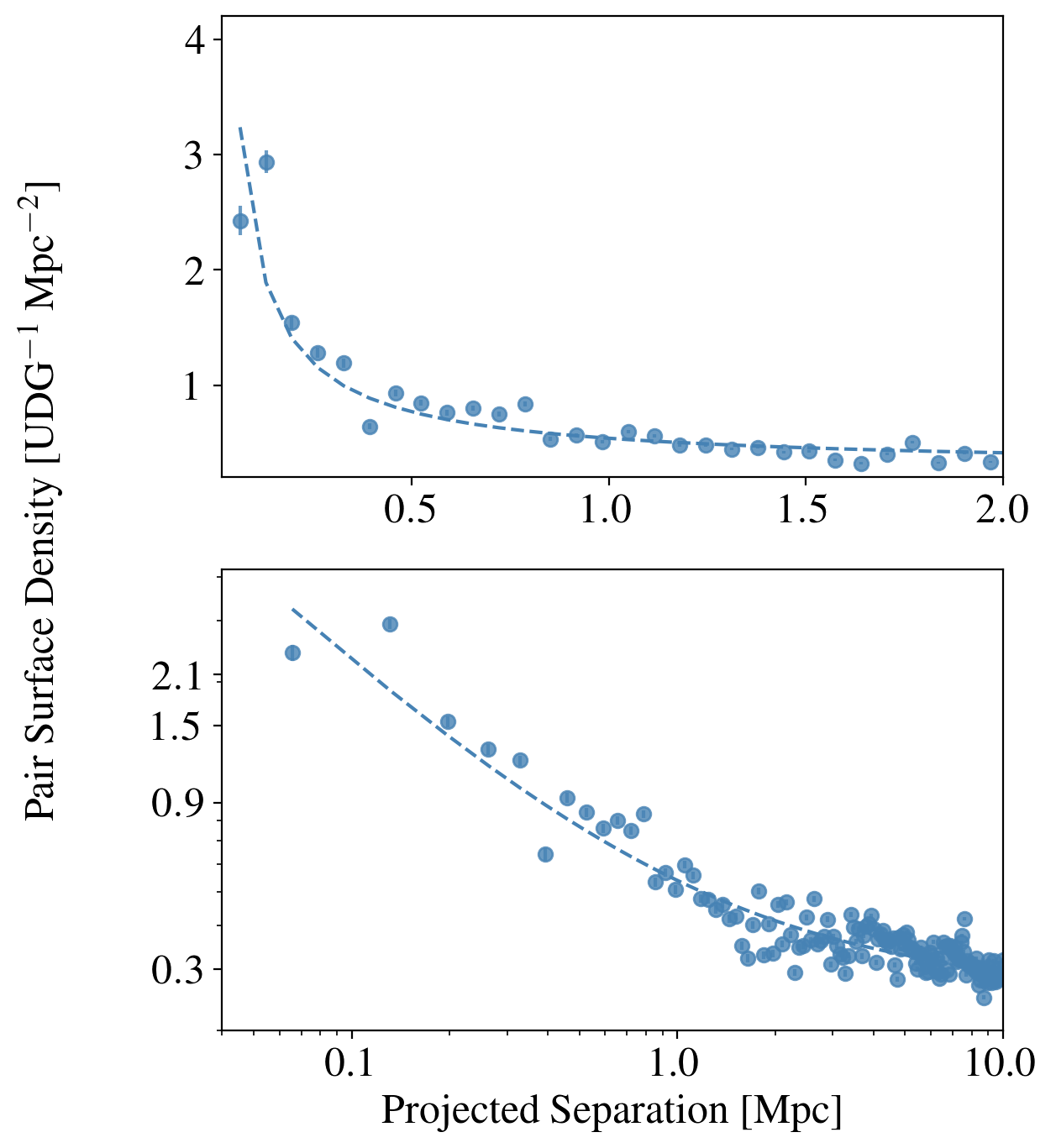}
    \caption{{UDG-MW pair surface density for the $4500 < cz/$(km s$^{-1}) < 5500$ slice per UDG. The pairs are defined by their projected separation. The upper panel shows the distribution in linear units, while the lower one in logarithmic units. The power law $+$ background model fit is performed in linear space but shown in both panels. Errorbars are plotted, but are mostly within the symbols themselves. We discuss the apparent underestimation of the uncertainties in the text.}}
    \label{fig:distribution}
\end{figure}

Armed with the distribution of pair separations, we present the corresponding pair distribution as a function of separation out to 10 Mpc (Figure \ref{fig:distribution}) using bins of 67 kpc width for the nearest redshift slice. We estimate
uncertainties using Poisson errors in the number of pairs in each bin and then propagate those through the calculation of the surface density. This is likely to be an underestimate of the uncertainties and we discuss the alternative approach below.
A simple power-law plus constant background fit to the surface density, $\Sigma = a r^k + b$, is also shown in Figure \ref{fig:distribution} and appears to be a reasonable approximation to the data (numerical values for the fit parameters are given in Table \ref{tab:fits} for each of the three redshift slices and color divided samples).

There are however two potential problems with the model fits and  their interpretation. First, as
we mentioned above, we expect incompleteness to set in at small separations. The comparison of the data and the fitted power-law
in Figure \ref{fig:distribution} suggests a possible decline in pair surface density in the innermost bin, but with our current binning scheme there is no resolution at radii within $r < 100$ kpc.
Whether the decline is observational, statistical, or physical is unclear. 

\begin{figure}[t]
    \centering
    \includegraphics[scale=0.5]{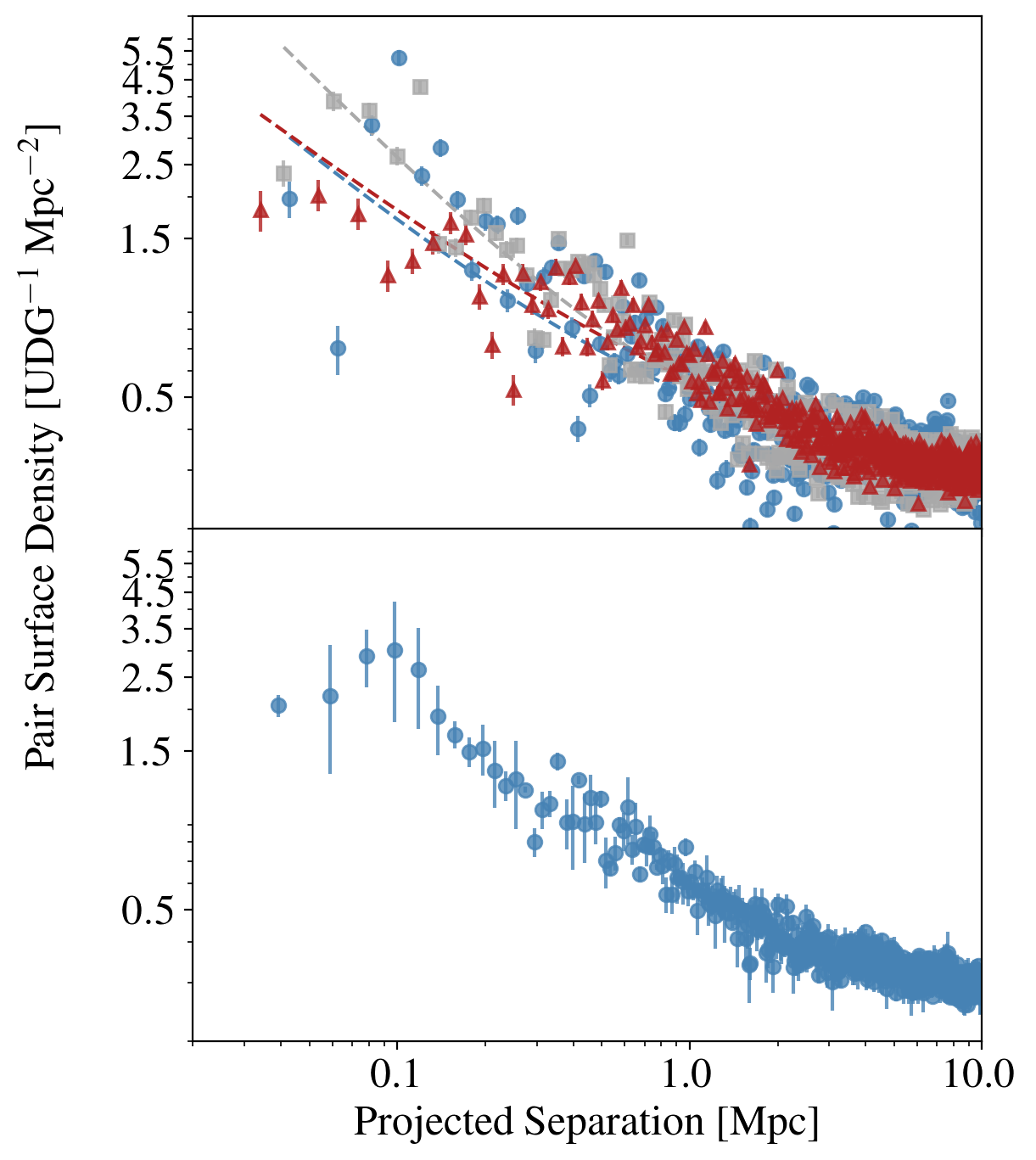}
    \caption{Comparison of pair separation distributions derived from the three redshift slices in the upper panel: $4500 < cz$/km s$^{-1} < 5500$ (blue circles); $5500 < cz$/km s$^{-1} < 6500$ (grey squares);
    $6500 < cz$/km s$^{-1} < 7500$ (red triangles). We have decreased the bin size to 20 kpc and overplot the fitted model curves for each of the three redshift slices. In the lower panel we show the average of the three slices with the uncertainties reflecting the error in the mean using standard deviation of those values rather than the Poisson uncertainties shown in the upper panel.}
    \label{fig:slice_comp}
\end{figure}

To further examine the behavior of the pair separation distribution at radii within 100 kpc, we reduce the bin width to 20 kpc with the understood sacrifice of lower statistical precision. The comparison of the pair surface densities derived from the three redshift slices is presented in Figure \ref{fig:slice_comp}. For this comparison, we scaled the three distributions to produce the same number density of background (uncorrelated) pairs. 
Although there is not an unambiguous systemic downturn among the distributions, a second potential problem becomes evident. The scatter among the measurements, even within one redshift slice, greatly exceeds the plotted statistical uncertainties, which are mostly smaller than the size of plotted symbols. We reevaluate the uncertainties using the scatter among the results from the three redshift slices and plot the mean and standard deviation of the mean  the lower panel of Figure \ref{fig:slice_comp}. This estimate of the uncertainties has the potential to be an overestimate as the surface density profiles among the slices may vary because the satellite samples in each slice are different in terms of luminosity and physical size. We expect the uncertainties to lie somewhere between those shown in Figure \ref{fig:distribution} and \ref{fig:slice_comp}, but closer to the latter.

In the lower panel of Figure \ref{fig:slice_comp} we find a  decline at the smallest radii (perhaps the innermost two or three bins), but the uncertainties are clearly larger than our previous estimates (excluding the innermost point, which is likely to have an actual uncertainty that is comparable to the other data within 100 kpc but just happened to contain measurements that exhibited little scatter). At the radii where this turnover might be detectable ($\sim 60$ kpc) some of our largest galaxy masks may be affecting the completeness. Examining the sample of confirmed UDG satellites discussed by \cite{karunakaran22}, we find no sign of such a turnover, but the numbers are small. On the other hand, if this turnover is real then it could signal an interesting physical effect. This topic is clearly an avenue that requires  further study with larger samples. We will discuss the possible effect of this uncertainty in our measurements below. As for why the Poisson statistics underestimate the true uncertainties, we suspect that it is related to the fact that pairs are not statistically independent (for example, a single UDG will be paired with each of the L$^*$ galaxies in a nearby galaxy group). 

\begin{deluxetable*}{lrrr}
\tablecaption{UDG-MWA Pair Separation Distribution Power-law Fit Parameters}
\label{tab:fits}
\tablewidth{0pt}
\tablehead{
\colhead{UDG Sample}&\colhead{$a$}&\colhead{$k$}&\colhead{$b$}
}
\startdata
All, 4500 $< cz/$km s$^{-1} < 5500$&0.28$\pm$0.02&$-0.87\pm0.07$&$0.26\pm 0.01$\\
All, 5500 $< cz/$km s$^{-1} < 6500$&0.21$\pm$0.01&$-0.96\pm0.05$&$0.18\pm 0.04$\\
All, 6500 $< cz/$km s$^{-1} < 7500$&0.32$\pm$0.01&$-0.70\pm0.04$&$0.21\pm 0.01$\\
\\
Red, 4500 $< cz/$km s$^{-1} < 5500$&0.39$\pm$0.03&$-0.86\pm0.07$&$0.23\pm 0.01$\\
Red, 5500 $< cz/$km s$^{-1} < 6500$&0.28$\pm$0.02&$-0.98\pm0.05$&$0.17\pm 0.01$\\
Red, 6500 $< cz/$km s$^{-1} < 7500$&0.38$\pm$0.02&$-0.73\pm0.05$&$0.20\pm 0.01$\\
\\
Blue, 4500 $< cz/$km s$^{-1} < 5500$&0.07$\pm$0.02&$-1.23\pm0.26$&$0.31\pm 0.01$\\
Blue, 5500 $< cz/$km s$^{-1} < 6500$&0.10$\pm$0.01&$-0.96\pm0.10$&$0.18\pm 0.04$\\
Blue, 6500 $< cz/$km s$^{-1} < 7500$&0.25$\pm$0.02&$-0.56\pm0.07$&$0.20\pm 0.01$\\
\enddata
\end{deluxetable*}

\begin{deluxetable*}{lr}
\tablecaption{Number of UDG Satellites per MWA}
\label{tab:numbers}
\tablewidth{0pt}
\tablehead{
\colhead{UDG Sample}&\colhead{$S_{\rm UDG}$}
}
\startdata
All, 4500 $< cz/$km s$^{-1} < 5500$&$0.53^{+0.10}_{-0.08}$\\
All, 5500 $< cz/$km s$^{-1} < 6500$&$0.46^{+0.05}_{-0.05}$\\
All, 6500 $< cz/$km s$^{-1} < 7500$&$0.25^{+0.02}_{-0.02}$\\
\\
Red, 4500 $< cz/$km s$^{-1} < 5500$&$0.42^{+0.08}_{-0.07}$\\
Red, 5500 $< cz/$km s$^{-1} < 6500$&$0.36^{+0.05}_{-0.04}$\\
Red, 6500 $< cz/$km s$^{-1} < 7500$&$0.16^{+0.02}_{-0.02}$\\
\\
Blue, 4500 $< cz/$km s$^{-1} < 5500$&$0.09^{+0.09}_{-0.04}$\\
Blue, 5500 $< cz/$km s$^{-1} < 6500$&$0.10^{+0.03}_{-0.02}$\\
Blue, 6500 $< cz/$km s$^{-1} < 7500$&$0.07^{+0.01}_{-0.01}$\\
\enddata
\end{deluxetable*}

To calculate the number of UDG satellites per MWA, $S_{\rm UDG}$, we invert our measurements. The distribution of separations remains the same, but the normalization changes. The one aspect we do not know is the number of MWAs over the survey area (we only searched for MWAs projected near UDGs). We can however, estimate the surface density of MWAs in the survey footprint using the measured background values, $b$, of our model fits. By multiplying this surface density and the survey area, $A$, of the full survey (20,000  deg$^2 $), we calculate the number of MWAs over the survey footprint ($\equiv A \cdot b$). Specifically, the number of UDGs associated with each MWA, for projected separations ranging from $r_{min}$ to $r_{max}$ is given by 
\begin{equation}
S_{\rm UDG} = \frac{\int_{r_{min}}^{r_{max}} 2 \pi  ar^kr dr \cdot N_{\rm UDG}}{A\cdot b},
\end{equation}
where $N_{\rm UDG}$ is the number of UDG candidates in the sample being considered (as opposed to the number of UDG satellites, $S_{\rm UDG}$) and $a$, $b$, and $k$ are the corresponding fit parameters for that sample.

We calculate $S_{\rm UDG}$ by integrating from 20 to 250 kpc. 
To estimate the uncertainties in $S_{\rm UDG}$, we evaluate the integral 1000 times, choosing different values for $a$, $b$, and $k$ from their respective error distributions, and use the resulting distribution of $S_{\rm UDG}$ to define  the 16th and 84th percentiles as the 1$\sigma$ uncertainty interval. The inner boundary of our integration represents a radius at which we are effectively within the MWA itself but in practice eliminating this cut does not increase the inferred $S_{\rm UDG}$ beyond the quoted uncertainty. The outer boundary represents the extent of the MWA halo, or virial radius. While our MW may have a somewhat smaller virial radius \citep[e.g.,][]{shen}, even decreasing the outer radius to 200 kpc reduces the inferred number of satellites only slightly below the quoted 1$\sigma$ lower bound quoted in Table \ref{tab:numbers}.
Finally, to address the possibility of a turnover in $S_{\rm UDG}$ at small radii,  if we integrate only from 70 kpc outward, where there is no hint of a turnover, $S_{\rm UDG}$ drops to 0.43 for the nearest redshift slice, which is a value that is only slightly more than a 1$\sigma$ decrease from that quoted. We expect this to be an overestimate of the  potential effect because we assumed in this test that there are absolutely no UDG satellites interior to 70 kpc even in projection. In fact, the region near the MWA may be a difficult one to interpret as there may be an additional contribution from UDG-like tidal dwarfs \citep{bennet18}.

\begin{deluxetable*}{lccc}
\tablecaption{Number of UDG Satellites per MWA in Luminosity Bins}
\label{tab:lf}
\tablewidth{0pt}
\tablehead{\colhead{Luminosity Range}&
\colhead{4500$< cz$/km s$^{-1} < 5500$}&\colhead{5500$< cz$/km s$^{-1} < 6500$}&\colhead{6500$< cz$/km s$^{-1} < 7500$}\\}
\startdata
$-17.0 < M_r < -16.0$&$0.02^{+0.16}_{-0.03}$&$0.02^{+0.01}_{-0.01}$&$0.07^{+0.02}_{-0.02}$\\
$-16.5 < M_r < -15.5$&$0.06^{+0.06}_{-0.03}$&$0.08^{+0.02}_{-0.02}$&$0.11^{+0.01}_{-0.01}$\\
$-16.0 < M_r < -15.0$&$0.14^{+0.03}_{-0.03}$&$0.24^{+0.04}_{-0.03}$&$0.16^{+0.02}_{-0.02}$\\
$-15.5 < M_r < -14.5$&$0.23^{+0.05}_{-0.04}$&$0.28^{+0.04}_{-0.04}$&$0.12^{+0.02}_{-0.01}$\\
$-15.0 < M_r < -14.0$&$0.26^{+0.06}_{-0.05}$&$0.18^{+0.03}_{-0.03}$&$0.06^{+0.01}_{-0.01}$\\
$-14.5 < M_r < -13.5$&$0.21^{+0.07}_{-0.05}$&$0.07^{+0.02}_{-0.02}$&$0.00^{+0.00}_{-0.00}$\\
\enddata
\end{deluxetable*}

\section{Results}
\label{sec:results}

\subsection{The UDG-MWA Pair Separation Distribution}

In Figures \ref{fig:distribution} and \ref{fig:slice_comp}, and Table \ref{tab:fits}, we present our measurements of the UDG-MWA pair separation distribution. 
The rise in the surface density toward smaller separations demonstrates that there does exist a significant population of UDGs that are physically correlated with MWAs. Additionally, the mean power law slope fit for three slices ($-0.84\pm0.06$) is in agreement with that for 'normal' satellites of giant galaxies \citep[$\sim -0.9$;][]{lorrimer}. UDGs do not appear to preferentially form or get destroyed in the environments near MWAs at different relative rates than do ``normal" satellites. This conclusion comes with the caveat that we have insufficient data to explore trends at radii smaller than about 60 kpc. Environment, at least broadly within the virial radius of MWAs, does not appear to be a significant factor in the evolution of the number of UDGs. This result is in concordance with a lack of any strong environmental signature in the approximately linear relation between the number of UDGs per halo and host halo mass extending from the most massive galaxy clusters down to MWAs \citep{karunakaran22,li}, which we will further confirm in \S \ref{sec:numbers}. 

Regarding UDG formation mechanisms, these
results indicate that UDGs, at least the population of
satellite UDGs, form primarily as part of the normal,
hierarchical universal dark matter superstructure \citep[e.g.,][]{dicintio,chan,jiang19,martin,wright}, rather than
through more specific channels like UDG formation through tidal interactions \citep{bennet18,jones21}, direct satellite collisions \citep{silk,shin}, or interaction with extremely dense environments \citep{yb,ss} that may best explain interesting individual
UDGs. Of course, even within the ``standard" model, formation for such a heterogeneous class of objects as UDGs may follow several formation pathways \citep{liao,sales20} and our measurement is insensitive to UDG satellites found at small separations.

\subsection{The Number of Satellite UDGs per MWA}
\label{sec:numbers}

Using the pair separation density profiles, we calculate and present the number of UDG satellites  for the typical MWA within projected radii between 20 and 250 kpc in Table \ref{tab:numbers}. Aside from the statistical errors that are quoted, these numbers are susceptible to various systematic uncertainties. First, the sample of UDGs is incomplete, as we will discuss in \S\ref{sec:colors}. Second, as we have discussed already, the appropriate limits on the integration  of the radial surface density profile are somewhat uncertain, which results in uncertainties that are comparable to the statistical ones. Lastly, we are working with projected rather than physical radii. 

\begin{figure}[ht]
    \centering
    \includegraphics[scale=0.5]{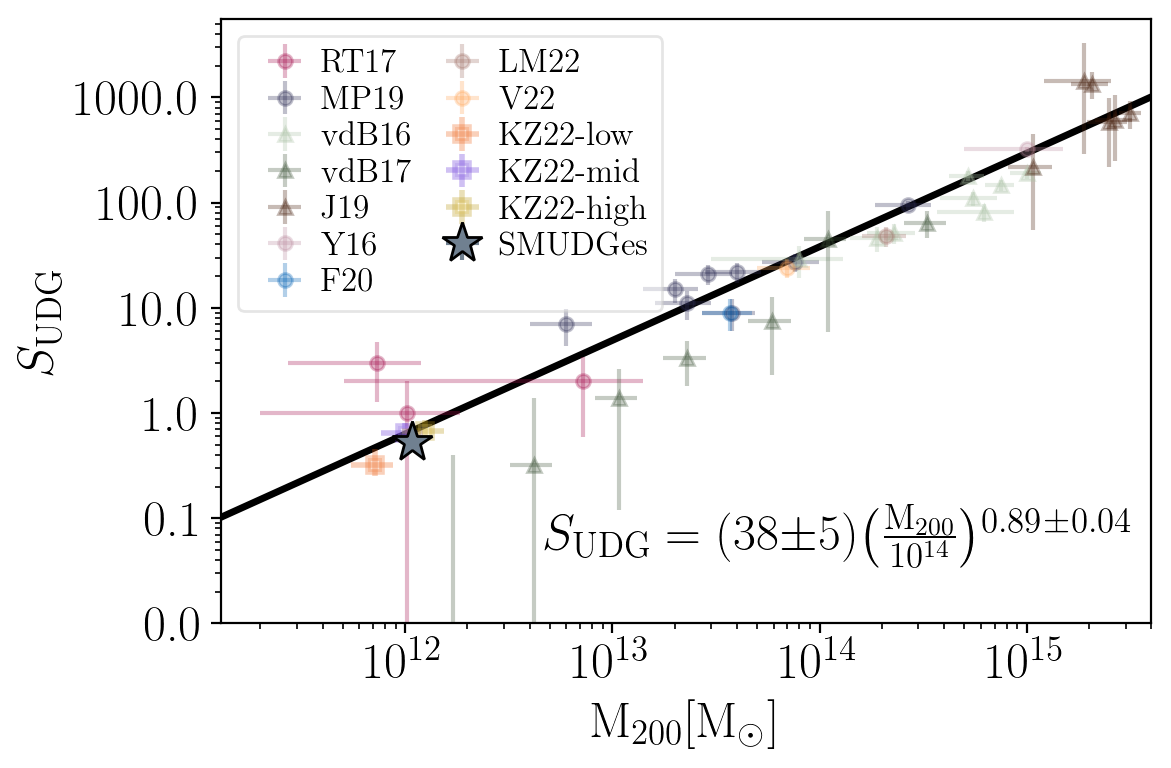}
    \caption{$S_{\rm UDG}$ vs. host halo mass. For our sample we adopt the MW mass estimate of \cite{shen}. The plot, the fitted relationship, and other measurements are adopted from \cite{karunakaran22} and the original measurements referenced in the legend \citep{rt17,mp19,vdb16,vdb17,j19,y16,f20,lm22,v22}.}
    \label{fig:halo}
\end{figure}

Focusing for now only on the full samples (not selecting by color), we find that the typical MWA contains less than one UDG satellite, within the range of UDG properties that we are sensitive to. 
We place this result in one context in Figure \ref{fig:halo} by comparing the number of UDG satellites per MWA, $S_{\rm UDG}$, with the numbers of UDG satellites measured in host halos of larger masses. The comparison is somewhat fraught because the data come from a variety of surveys that have different selection criteria. Such differences tend to be of order unity and are obscured by the large parameter range covered in the Figure, but they need to be carefully addressed if one is interested in modest deviations from a linear slope in the overall relation. Nevertheless, we confirm the qualitative conclusion of previous studies \citep{karunakaran22, li} that $S_{\rm UDG}$ for MWAs is approximately consistent with a linear extrapolation of the relation established using halos of larger total mass.

The near linearity of this relation over more than 3 orders of magnitude in mass might appear to challenge models where UDG satellites are a hybrid population, for example those where a significant fraction of UDG satellites are born as such and the remainder consists of galaxies transformed by the environment \citep{liao,sales20}. However, at least in terms of the number of UDG satellites, the \cite{liao} study is consistent with our finding --- predicting $\sim$ 1 UDG satellite per MWA. Nevertheless, 
there is likely to be a fine tuning challenge if simulations showing that only a small fraction of UDGs that fall into clusters survive \citep[$\sim$20\%;][]{jiang19} are correct. This challenge could become acute when a precise slope is empirically determined.  
A version of Figure \ref{fig:halo} redone with homogeneous data and selection is critical to further confrontation to the models.

\subsection{UDG Satellite Colors}
\label{sec:colors}

\begin{figure}[ht]
    \centering
    \includegraphics[scale=0.5]{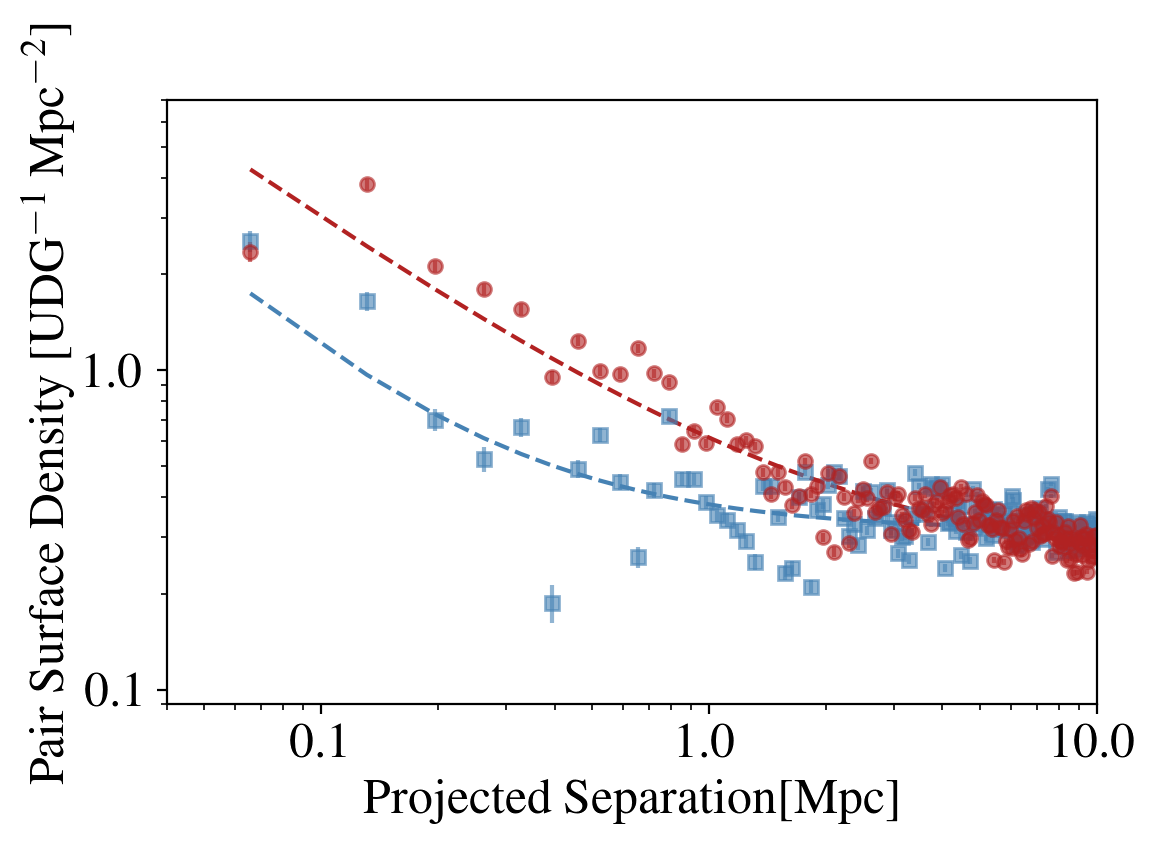}
    \caption{{Pair surface density as a function of UDG color. UDG-MW pairs that include a UDG classified as red are shown in the red circles, while those classified as blue are shown as blue squares.  Dashed lines are power law plus flat background fits to the data. }}
    \label{fig:colors}
\end{figure}

We now examine the behavior of red and blue UDGs separately. In Figure \ref{fig:colors} we present the pair separation distribution for red and blue UDGs in the redshift slice spanning $4500 < cz/$(km s$^{-1}) < 5500$. The results for the other slices are similar.
While the surface density
rise at small separations is present in both populations, confirming the existence of both red and blue UDG satellite populations, 
it is dominated by red UDGs, indicating that the majority of UDG satellites of MWAs have not recently been forming stars at a significant rate. 
Roughly 80\% of the UDG satellites we find are red.
As
such, and in contrast to our conclusion regarding the number of UDGs, there is a strong environmental signature in the stellar populations 
of UDG satellites. Interestingly,
however, there are some star-forming UDGs even at
small (projected) separations, suggesting that 
whatever environmental quenching there may be is either not rapid or entirely efficient. This result follows on the suggestion from \cite{karunakaran21} that the overall satellite populations of galaxies indicate that quenching may be overestimated in current simulations. We close by noting that the divergence by color in the populations is evident even at large radii ($>$ 1 Mpc), well outside the virial radii of MWAs, much like it is in the general galaxy population surrounding galaxy clusters \citep{lewis,gomez}. As such, it suggests that much like for more massive galaxies, a full understanding of the environmental effects will be challenging to reach and must involve pre-processing 
\citep{zm,mcgee,delucia} that occurs prior to the galaxy's arrival in its current environment. 

\subsection{UDG Satellite Luminosity Function}

\begin{figure}[ht]
    \centering
    \includegraphics[scale=0.5]{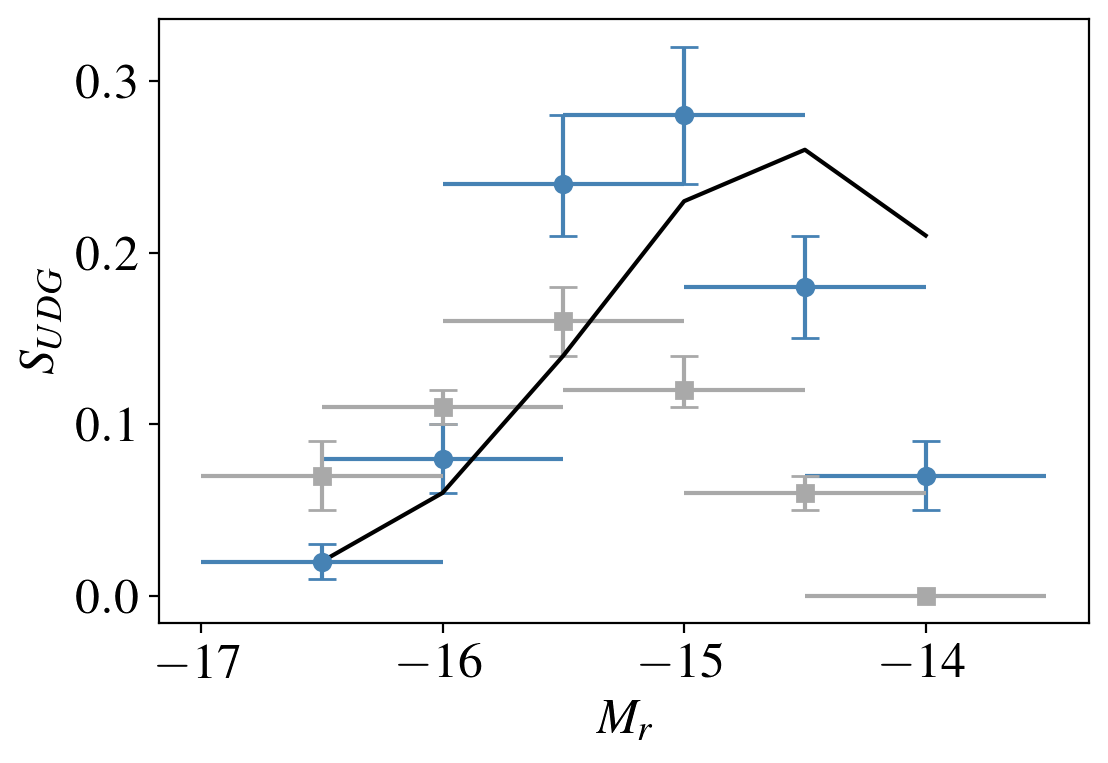}
    \caption{{UDG satellite luminosity function. We present the number of UDG satellites within 1 magnitude bins as derived from our three redshift slices. The results for the nearest slice ($4500 < cz/$(km s$^{-1}) < 5500$) are represented by the black line, while those of the other two slices are represented by blue circles and gray squares, with the squares representing the farthest of the three slices. Horizontal bars represent the bin widths, while vertical error bars are statistical uncertainties. Units for $S_{\mathrm{UDG}}$, are number per MWA per magnitude.}}
    \label{fig:lf}
\end{figure}

For each of the three redshift slices, we present the number of UDG satellites per magnitude over the range of magnitudes to which we are sensitive in Table \ref{tab:lf} and Figure \ref{fig:lf}. Together these provide both a selection-uncorrected view of the UDG satellite luminosity function and the associated uncertainties. 

The results among the three redshift slices agree well down to M$_r \sim -15$ and then begin to diverge. That divergence is systematic in that the decline in numbers with fainter luminosity begins at brighter luminosity with the most distant redshift slice and continues in sequence up until the nearest slice. 
This behavior is as expected because we are less sensitive both to fainter and smaller UDG satellites at larger distances. Even in the intermediate redshift slice, the SMUDGes angular selection criterion already excludes UDGs with physical effective radii of less than about 2 kpc. The turnover in the luminosity function obtained from the nearest redshift slice suggests that it too is incomplete below M$_r \sim - 15$. If 
so, then this means that the total satellite numbers we provide in \S\ref{sec:numbers} are, to some degree, underestimates of the full number of UDG satellites. However, there are reasons to believe that the UDG population does not extend in large numbers to fainter surface brightnesses than those captured by SMUDGes \citep{smudges3}.

To place these numbers in context, we compare our luminosity function for UDG satellites with the satellite luminosity functions of four nearby, extremely well-studied MWAs \citep[M 84, M 91, M 101, and Cen A;][and references therein]{bennet19} in Figure \ref{fig:lf2}. We show the cumulative satellite luminosity function for the combined set of nearby MWAs and that for our UDG satellites. We have corrected our values of M$_r$ to M$_{\rm V}$ using a mean value of $g-r$ of 0.6 for UDGs and a correction \citep{fukugita} from V to $g$ of 0.2 mag for early type galaxies (80\% of UDG satellites are non star forming; \S\ref{sec:colors}).
We expect that the published luminosity functions for the local MWAs include any UDG satellites that are there because those studies used deep, wide field observations intended to reach both faint and low luminosity systems. For $-16 < {\rm M}_{\rm V} < -14$, we find that UDG satellites are $\sim$ 10\% of the satellite population. 
We conclude that the UDG satellite population at a given luminosity, for M$_{\rm V} < -14$, is well sub-dominant and there is no significant, lurking population of large, low surface brightness satellites at these luminosities.

\subsection{UDG Satellite Mass Function}

\begin{figure}[ht]
    \centering
    \includegraphics[scale=0.5]{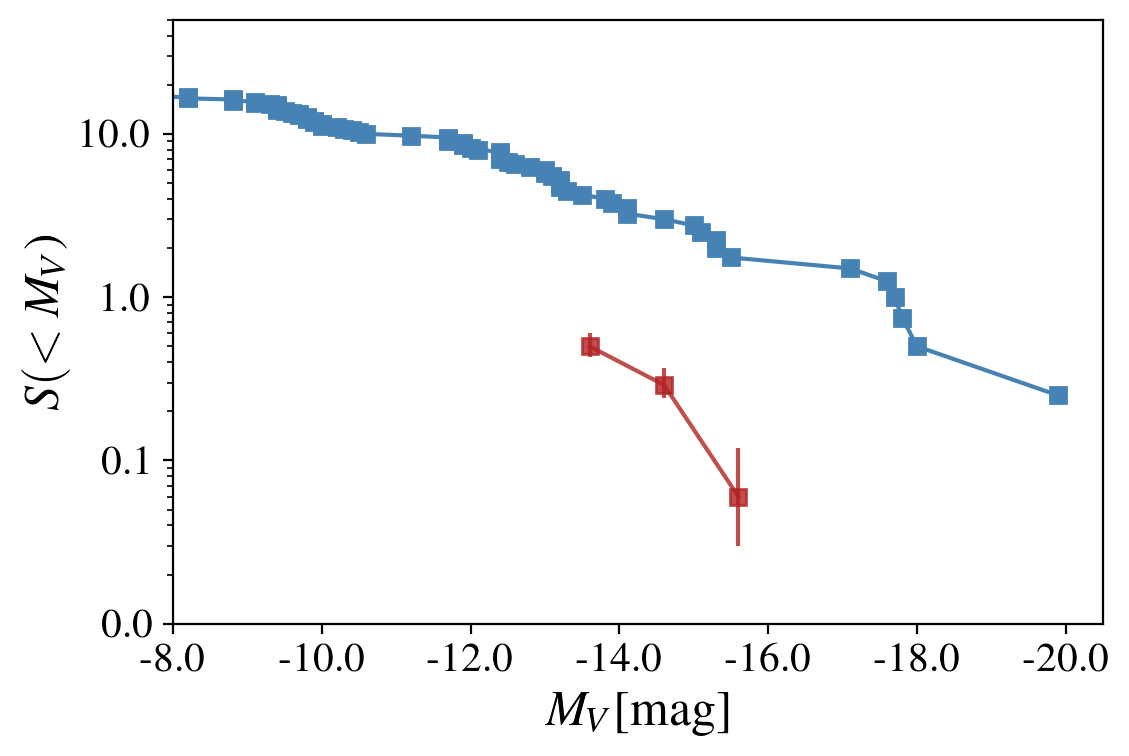}
    \caption{{Comparison of cumulative satellite luminosity function for local MWAs (data from Bennet et al. 2019 and references thererin, see text) in blue and our mean measurements for UDG satellites of MWA in red. At a given luminosity, over the range of our measurements, UDG satellites are typically about $\sim$10\% of the total number of satellites.}}
    \label{fig:lf2}
\end{figure}

A mass function measurement would be ideal for a direct comparison to models. Although cosmological simulations do produce UDGs \citep{tremmel, wright}, we are always at the mercy of assumptions in the baryonic sub-grid physics if we can only compare the luminous properties of galaxies. A check on those assumptions would be to have both the luminosities and masses of UDGs (or at least internal kinematics).
As we mentioned previously, the total mass-to-light ($M/L_{total}$) ratios of UDGs are likely to be significantly larger than those of comparably massive galaxies. 
At the limit of our current understanding of UDGs, they appear to have $M/L_{total}$ that is at least an order of magnitude larger  \citep{vdk19df44}, with perhaps some unusual exceptions \citep{vdk22}. If indeed $M/L_{total}$ for UDG satellites is a factor of 10 larger than for non-UDG satellites of similar luminosity, then we 
should slide the UDG LF in Figure \ref{fig:lf2} to the right by 2.5 magnitudes to appropriately compare the numbers of similarly {\sl massive} satellites. At this point, the UDG satellites would still be subdominant in number, but now only by a factor of a few rather than an order of magnitude. As such, they could play a significant role in the satellite/subhalo accounting at LMC-like masses.

To explore this topic a bit further, we  estimate the total mass of these low mass galaxies using only photometry \citep{zb}. In this approach, a scaling relation is used to recover the velocity dispersion at the effective radius and, therefore, an estimate of the enclosed mass within this radius. By assuming an NFW density profile \citep{nfw}, we then determine which model produces the measured enclosed mass at $r_e$. The method was used by \cite{zb} to explore the stellar mass-halo mass relation and by \cite{z22} to study the relation between globular cluster populations and total mass. We use the relation to isolate UDGs with masses comparable to or larger than that of the LMC \citep[log $(M_h/M_{\odot}) = 11.14$;][]{erkal}. We present results for  pair separation involving UDGs inferred to have $10.9 < \log (M_h/M_{\odot}) < 12$. As a caution, we note that the scaling relation has not been fully vetted to apply to UDGs because of the paucity of spectroscopic data for UDGs. Where comparison is possible, the results are in acceptable agreement and provide masses within a factor of a few, which is comparable to the overall precision limit of the method and within the range of our order of magnitude mass selection bin. Further discussion of the use of this approach for UDGs will be presented in \cite{smudges4}.

We find results from the three slices that are consistent ($0.08^{+0.08}_{-0.04}, 0.06^{+0.04}_{-0.03},$ and $0.07\pm0.1$, for the lowest to highest redshift slices respectively) and that on average suggest that 
the number of UDG satellites with LMC-like or larger masses per MWA is $0.07^{+0.02}_{-0.01}$, or alternatively, corresponding to $\sim$ 13\% of our deepest satellite sample. Roughly 1 in 14 MWAs have a UDG satellite that is of comparable mass to the LMC.
Comparing that result to the calculation based on standard $\Lambda$CDM that $\sim$ 40\% of $10^{12}$ M$_{200}$ halos should host something nearly as massive as the LMC \citep{wang} suggests that  $\sim$ 18\% of this population may fall in the ultra-diffuse class. 

\section{Conclusions}

We present a correlation analysis between UDG candidates from the SMUDGes catalog \citep{smudges,smudges3,smudges4} and Milky Way analogs (MWAs) drawn from the SIMBAD database \citep{simbad} from which we identify a population of UDGs that are physically associated with MWAs. We find the following:

\smallskip
\noindent
$\bullet$
A population of UDG satellites exists that surrounds MWAs. The distribution of those satellites (projected surface density $\propto r^{-0.84\pm0.06}$) is entirely consistent in character with that of normal satellite galaxies. We conclude that the processes by which most of these UDG satellites form are related to how low mass galaxies form in general. We exclude exotic formation mechanisms for UDG satellites as a primary formation channel.
A consistent conclusion was reached in a recent study of an entirely different population of UDGs \citep{jones23}.

\smallskip
\noindent
$\bullet$
On average, each MWA has $\sim 0.5\pm0.1$ UDG satellites at projected radii between 20 and 250 kpc and $-17 < {\rm M}_r < -13.5$. 

\smallskip
\noindent
$\bullet$
We compare our measurement of the number of UDG satellites per MWA to published measurements of the number of UDG satellites in hosts of different masses.
We confirm previous findings that the number of UDG satellites of MWAs is consistent with a nearly linear trend between the number of UDG satellites and total halo mass \citep{karunakaran22,li}. We interpret this finding as providing further evidence against specific, UDG formation scenarios that are unconnected with the general formation path of low mass galaxies. 

\smallskip
\noindent
$\bullet$
We find that red UDGs are far more tightly clustered around MWAs than blue UDGs and that red UDGs comprise $\sim$ 80\% of the UDG satellite population of MWAs out to 250 kpc (where blue is defined as being more than 0.1 mag bluer than the red sequence in $g-r$ vs. M$_r$). Although environmental quenching is likely involved, we note that, as with normal galaxies near galaxy clusters \citep{lewis,gomez}, the color changes happen well outside the virial radius and the trend likely results from  a far more complex history \citep[e.g.,][]{delucia} than that of simple quenching scenarios.

\smallskip
\noindent
$\bullet$
We find that for $-17 < {\rm M}_r < -13.5$ UDG satellites are $\sim$ 10\% of the total satellite population down to a similar magnitude limit. However, we note that UDGs have been shown, in the limited number of cases studied, to be strongly dark matter dominated and may therefore represent a larger fraction of satellites down to a correspondingly larger total mass limit. In support of this claim we estimate halo masses using the \cite{zb} methodology and conclude that UDG satellites may comprise $\sim$18\% of the satellites with halo masses of at least half the mass of the LMC.

In summary, UDG satellites appear to be directly connected to the overall satellite population in a manner that suggests that there is not a distinct, separate formation channel. They are a minority, but still significant fraction of the satellite populations of Milky Way analogs and should be included in discussions involving satellite galaxy populations.

\begin{acknowledgments}

The authors acknowledge financial support from NSF AST-1713841 and AST-2006785 for SMUDGes.  An allocation of computer time from the UA Research Computing High Performance Computing (HPC) at the University of Arizona and the prompt assistance of the associated computer support group is also gratefully acknowledged. AK acknowledges financial support from the grant CEX2021-001131-S funded by MCIN/AEI/ 10.13039/501100011033 and from the grant POSTDOC\_21\_00845 funded by the Economic Transformation, Industry, Knowledge and Universities Council of the Regional Government of Andalusia.

\end{acknowledgments}

\software{
Astropy              \citep{astropy1, astropy2},
astroquery           \citep{astroquery},
galpy,               \citep{galpy},
Matplotlib           \citep{matplotlib},
NumPy                \citep{numpy},
pandas               \citep{pandas},
SciPy                \citep{scipy1, scipy2},
}

\bibliography{references.bib}{}
\bibliographystyle{aasjournal}

\end{document}